\documentclass[preprintnumbers,amsmath,amssymb,floatfix,12pt,prd,superscriptaddress,nofootinbib]{revtex4}
\usepackage{graphicx}
\usepackage{epsfig}
\usepackage{bm}
\usepackage{amsfonts}

\def\ben{\begin{equation}}
\def\een{\end{equation}}

 \def\bd{\begin{document}} \def\ed{\end{document}}
\def\ds{\documentstyle} \let\fr=\frac \let\bl=\bigl \let\br=\bigr
\let\Br=\Bigr \let\Bl=\Bigl
\let\bm=\bibitem
\let\na=\nabla
\let\pa=\partial \let\ov=\overline
\newcommand{\be}{\begin{equation}}
\newcommand{\ee}{\end{equation}}
\def\ba{\begin{array}}
\def\ea{\end{array}}
\def\ft#1#2{{\textstyle{\frac{\scriptstyle #1}{\scriptstyle #2} } }}
\def\fft#1#2{{\frac{#1}{#2}}}
\def\del{\partial}
\def\vp{\varphi}
\def\sst#1{{\scriptscriptstyle #1}}
\def\oneone{\rlap 1\mkern4mu{\rm l}}
\def\td{\tilde}
\def\wtd{\widetilde}
\def\ie{{\it i.e.\ }}
\def\dalemb#1#2{{\vbox{\hrule height .#2pt
        \hbox{\vrule width.#2pt height#1pt \kern#1pt
                \vrule width.#2pt}
        \hrule height.#2pt}}}
\def\square{\mathord{\dalemb{6.8}{7}\hbox{\hskip1pt}}}
\newcommand{\ho}[1]{$\, ^{#1}$}
\newcommand{\hoch}[1]{$\, ^{#1}$}
\newcommand{\bea}{\setlength\arraycolsep{2pt} \begin{eqnarray}}
\newcommand{\eea}{\end{eqnarray}}
\newcommand{\ra}{\rightarrow}
\newcommand{\lra}{\longrightarrow}
\newcommand{\Lra}{\Leftrightarrow}
\newcommand{\bp}{\tilde \beta^\prime}
\newcommand{\tr}{{\rm tr} }
\newcommand{\Tr}{{\rm Tr} }
\def\0{{\sst{(0)}}}
\def\1{{\sst{(1)}}}
\def\2{{\sst{(2)}}}
\def\3{{\sst{(3)}}}
\def\4{{\sst{(4)}}}
\def\5{{\sst{(5)}}}
\def\6{{\sst{(6)}}}
\def\7{{\sst{(7)}}}
\def\8{{\sst{(8)}}}
\def\m{{\sst{(m)}}}
\def\n{{\sst{(n)}}}
\def\cA{{{\cal A}}}
\def\cB{{{\cal B}}}
\def\cF{{{\cal F}}}
\def\cG{{{\cal G}}}
\def\cH{{{\cal H}}}
\def\tV{\widetilde V}
\def\tW{\widetilde W}
\def\tH{\widetilde H}
\def\tE{\widetilde E}
\def\tF{\widetilde F}
\def\tA{\widetilde A}
\def\im{{{\rm i}}}
\def\tY{{{\wtd Y}}}
\def\ep{{\epsilon}}
\def\vep{{\varepsilon}}
\def\bD{{{\bar D}}}
\def\R{{{\mathbb R}}}
\def\C{{{\mathbb C}}}
\def\H{{{\mathbb H}}}
\def\CP{{{\mathbb C}{\mathbb P}}}
\def\RP{{{\mathbb R}{\mathbb P}}}
\def\Z{{{\mathbb Z}}}
\def\bA{{{\mathbb A}}}
\def\bB{{{\mathbb B}}}
\def\bC{{{\mathbb C}}}
\def\bD{{{\mathbb D}}}
\def\bE{{{\mathbb E}}}
\def\bZ{{{\mathbb Z}}}
\def\Re{{{\frak{Re}}}}
\def\Im{{{\frak{Im}}}}
\def\cosec{{\,\hbox{cosec}\,}}
\def\Gm{{\Gamma_{\!\! -}}}
\def\Gp{{\Gamma_{\!\! +}}}
\def\stan{{standard }}
\def\nonstan{{supernumerary }}
\def\p{{\partial}}
\def\kdel#1{{\fft{\del}{\del#1}}}

\def\bog{{Bogomolny }}
\def\om{{\omega}}

\newcommand{\nnr}{\nonumber \\}
\newcommand{\pd}{\partial}
\newcommand{\ud}{\textrm{d}}
\newcommand{\dTH}{T^{\prime \, 0}_\textrm{H}}
\newcommand{\dOi}{\Omega^{\prime \, 0}_i}
\newcommand{\bx}{{\bf x}}
\begin{document}

\title{Caustic singularity
 in Ho$\check {\textbf{r}}$ava-Lifshitz gravity}
\author{\textbf{M.R.Setare}}
\email{rezakord@mail.ipm.ir} \affiliation{Department of Campus of Bijar , University of Kurdistan,
Bijar, IRAN.}

\author{\textbf{D. Momeni}}
\email{d.momeni@yahoo.com} \affiliation{Department of Physics,
Faculty of  Sciences, Tarbiat Moa'llem University, Tehran, Iran}

\begin{abstract}
\vspace*{1.5cm} \centerline{\bf Abstract} \vspace*{1cm}

In this note we searched
 for a family of solutions with Caustic singularity in non
relativistic-renormalizable Ho$\check {\textbf{r}}$ava-Lifshitz (HL)
theory without the general covariant. We show that in infrared
 (IR) limit and with a deviation from $\lambda=1$ we have no caustic
 singularity. Also in ultraviolet (UV) regime and for Ricci flat 3-dimensional ($3d$) spaces
and codimension 1 and for $\lambda\neq1$ the non linear
 terms should help bouncing this kind of most dangerous would be
 caustics. But if $3d$ curvature does not vanish, higher curvature terms
 do help caustics even in codimension one. Thus the arguments in [JCAP 0909:005,2009] are
 satisfied correctly.
\end{abstract}

\maketitle

\newpage

\section{Introduction}
 HL theory \cite{hor1,hor2}is a non relativistic foliation preserving
 homeomorphism invariant theory which is stochastic quantized in UV
 region and in IR limit it mimics the general relativity (GR) with a dark
 matter\cite{Mukohyama-CQG,PRD}. In the other word the gauge
 symmetries of the system are foliation-preserving diffeomorphisms of
 spacetime. Also the higher curvature terms in the action can
 be treated as a generalized modified gravity. They lead to the regular
 solutions in the UV regime and also make the flatness problem
 milder\cite{Calcagni}. The anisotropic scaling of this model in $z=3$
 critical point  solves the horizon problem and also
 describes the scale invariant cosmological perturbations without any
 need to the inflation\cite{Kiritsis,JCAP}. As a field theoretic
 model, HL theory is power counting renormalizable in spite of the GR
which is not power counting. This is one of the most difficulties of
the quantum gravity. If the critical exponent $z$ is fixed in
 $z=3$, the amplitude of quantum fluctuations of the scalar field does
 not change as the energy scale of the system changes. In this case
 the non linear interactions of graviton are power counting
 renormalizable and for such values of $z>3$ is super
 renormalizable\cite{Cai}. The power counting renormalizability is
 achieved by violating the Lorentz invariance with working by
 anisotropic scaling of time and spatial coordinates. Unlike the GR in
 HL theory if we want to preserve the foliation preserving
 diffeomorphism invariance theory must be parity invariance. The original HL model was parity violating.
  It was subsequent
work by Sotiriou et al.\cite{Sotiriou} that showed that a parity
respecting variant of the original HL model could be constructed. It
is not
 the unique option. More recently Horava and Melby-Thompson proposed a new version
\cite{HMT}. In this
 version the  extended gauge symmetry eliminates the scalar graviton
 and consequently limited the value of the coupling constant
  $\lambda$. Outset we show that if we take a general plane
  symmetric background for propagation of fields, the IR limit of
  equation of motions register the familiar $1+1$ dimensional wave
  equation. Secondly this equation designates a square term proportions
  to the Cosmological constant. Thus we can treat $\Lambda$ in IR limit
  as a key for repulsive attraction. Further we show that there is no possibility for
   Caustic singularity formation, i.e. a singular  3-dimensional
Ricci flat extrinsic
  curvature solution not by higher order terms nor in IR limit. But
  in UV if  $3d$ curvature does not vanish,higher curvature terms
 do help caustics even in codimension one.

  \section{IR limit of HL gravity}
  Following the Mukohyama\cite{Mukohyama-CQG}, at now we know that the
  HL theory mimics GR plus dark matter\cite{Caustic}. Also between four
  different versions of this theory only one  with projectable
  condition and without detailed balance (or with a small deviation
  from this principle) is guaranteed. The general action for HL theory
  with these conditions reads as

  \begin{eqnarray}
  I_{IR}=I_{Kin}+I_{z=1}+I_{z=0}=\frac{M_{Pl}^{2}}{2}\int
  Ndt\sqrt{g}d^{3}\overrightarrow{x}(K^{ij}K_{ij}-\lambda
  K^{2}+R-2\Lambda)
  \end{eqnarray}

  with an ADM metric \cite{ADM}in 3+1 decomposition formalism
  \begin{eqnarray}
  ds^{2}=-N(t)^{2}dt^{2}+g_{ij}dx^{i}dx^{j}
  \end{eqnarray}
   In order to
  host instability,thus $\lambda$ must be either larger than 1 or
  smaller than 1/3. We know that $\lambda$ runs from $+\infty$ in the UV to $1+0$ in
  the IR. According to a phenomenological constraint on properties of
  the renormalization group(RG), $\lambda$ must be sufficiently close to 1 at low
  energy,while $\lambda-1$ can be of $O(1)$at larger at high energy.
   In IR limit or macroscopic objects neglecting from the higher spatial
  derivatives terms $I_{z=3}, I_{z=2}$  in the action (1) is the
  single  option. The lapse function $N(t)$ is required to be independent of
  spatial coordinates by the profitability condition. Hence by a
  space-independent time parameterizations we set the lapse to unity.
  \subsection{Equations of motion}
  By variation of the action (1) in the absence of any matter field
  with respect to the lapse function N(t), we obtain the Hamiltonian
  constraint which is global constraint and not the local
  constraint one's
  \begin{eqnarray}
  H_{g}=-\frac{\delta I_{g}}{\delta N}=\int d^{3}\overrightarrow{x}
  \mathcal{H}_{g}=\frac{M_{Pl}^2}{2}\sqrt{g}(K^{ij}p_{ij}-\Lambda-R-I_
{z1})\\\nonumber
  p_{ij}=K_{ij}-\lambda K g_{ij}
  \end{eqnarray}
   For shift variation i.e $N^{i}(t,\overrightarrow{x})$ we
  have the next divergence's like equation for reduced extrinsic
  curvature tensor $p_{ij}$
  \begin{eqnarray}
  -M_{Pl}^2\sqrt{g}\nabla^{i}p_{ij}=0
  \end{eqnarray}

  and finally the $3d$ metric must obeys from the following  equation of motion

  \begin{eqnarray}
  \epsilon_{gij}=g_{ik}g_{jl}(\frac{2}{N\sqrt{g}}\frac{\delta
  I_{g}}{\delta
  g_{kl}})=M_{Pl}^2[-\frac{1}{N}(\partial_{t}-N^{k}\nabla_{k})p_{ij}
\\\nonumber
  +\frac{1}{N}(p_{ik}\nabla_{j}N^{k}+p_{jk}\nabla_{i}N^{k})-Kp_{ij}+2K^{k}_
{i}p_{kj}+\frac{1}{2}g_{ij}K^{kl}p_{kl}\\\nonumber
  +\frac{1}{2}\Lambda g_{ij}-G_{ij}]+\epsilon_{z>1,ij}=0
  \end{eqnarray}

  \section{Metric }
We shall restrict ourselves to situations where the space time has
plane symmetry. We adopted a coordinates system $(t,x,y,z)$ with
plane symmetry such
  that for them the projectable case satisfied by setting $N(t)=1$ and
  the 3 spatial metric $g_{ij}$ be in the form
  \begin{eqnarray}
  g_{ij}=diag(1,e^{u(x,t)},e^{u(x,t)})
  \end{eqnarray}
  and the coordinates $i,j={x,y,z}$.
 That is, the problem is invariant under
transformations of the form
\begin{eqnarray}\nonumber
y\rightarrow y+a\\\nonumber z\rightarrow z+b\\\nonumber ,\\\nonumber
y\rightarrow y\cos(\theta)+z\sin(\theta)\\\nonumber z\rightarrow
z\cos(\theta)-y\sin(\theta)
\end{eqnarray}
The  four dimensional metric of a spacetime admitting plane symmetry
in general may be written as \cite{Taub}(we take $c=1$)
\begin{eqnarray}\nonumber
ds^2=-e^{2F}dt^2+e^{2G}dx^2+e^{2H}(dy^2+dz^2)
\end{eqnarray}
Where $F,G$ and $H$ are functions  of $x$ and $t$ alone.  Static and
non static exact solutions for HL gravity has been investigated by
the authors in Ref. \cite{IJMPD}.

   For future uses we write the Ricci
scalar for $3d$ part of (6) as
  \footnote{$f_{a}=\frac{\partial f}{\partial
x^{a}},\dot{u}=\frac{\partial u}{\partial t}$}
 \begin{eqnarray}
 R=2u_{xx}+\frac{3}{2}u_{x}^2
  \end{eqnarray}

  This form of metric is planar and
  stationary thus it is a good example for testing the Mukohyama's
  idea about the no existence of the vacuum caustic singularities in
  HL theory. We are searching for at leat one $3d$ Ricci flat$(R=0)$
  solution for metric ansatz (6) in such a way that it's extrinsic
  curvature $K=u_{t}$ diverges. The general form of Ricci flat metric's function may be written as
  the following
 \begin{eqnarray}
 u(x,t)=\frac{4}{3}\log(\frac{3}{4}(f_{1}(t)x+f_{2}(t))
  \end{eqnarray}
Or the following equivalent form
\begin{eqnarray}
 u(x,t)=\frac{4}{3}\log(f_{1}(t))+\frac{4}{3}\log(x+f_{3}(t))+c
  \end{eqnarray}
  Where in it $c=(4/3) \ln(3/4)$. If  we define $f_4(t)$ by $f_1(t) =
  (4/3)f_4(t)$
then (9) becomes
\begin{eqnarray}\nonumber
 u(x,t)=\frac{4}{3}\log(f_{4}(t))+\frac{4}{3}\log(x+f_{3}(t))
  \end{eqnarray}
  So without any loss of generality, the constant c can be eliminated
completely. The unknown functions $f_{i(=1,3)}(t)$ must be
determined via substituting in field equations.

\section {Exact solution for IR limit}
In this section we search for all exact solutions for field
equations (4), (5) with metric ansatz (6). We divided exact
solutions to 2 different class: One with (R=0) and another without
this constraint. In first case we will show that the Caustic
singularity avoids in a natural sence. But in the another case the
field equations reduces to a (1+1) dimensional wave equation with an
effective sound speed with a wide class of solutions. In this class
we can treat $\Lambda$ as a repulsive force.
  \subsection{Ricci flat solution for field equations in IR limit}
   The field equations (4), (5)after variation action (1) for lapse, shift and
   2 independent (but symmetric) components of the metric lead to the
   next set of partial differential equations
  \begin{eqnarray}
  -(1-2\lambda)\dot{u}^{2}+\Lambda+2u_{xx}+\frac{3}{2}u_{x}^{2}=0\\
  \lambda\ddot{u}+\frac{1}{4}(1+2\lambda)\dot{u}^{2}+\frac{1}{4}u_{x}^{2}
+\frac{1}{2}\Lambda=0\\
  -(1-2\lambda)\ddot{u}+\Lambda+u_{xx}+\frac{u_{x}^{2}}{2}=0
  \end{eqnarray}
We substitute (9) in (10)-(12). In this case (10) gives us
\begin{eqnarray}
\dot{u}=\pm\sqrt{\frac{\Lambda}{1-2\lambda}}
\end{eqnarray}
Since $\Lambda>0$ to avoidance from a pure imaginary solution we
must have
\begin{eqnarray}
\lambda\leq\frac{1}{2}
\end{eqnarray}
If the equality sign holds we have a singularity in $u_{t}$ and
consequently from it $K=\infty$. But we remove this special case
since we want to keep the value of $\lambda$ near 1 and also as we
will show in next lines we must set $\Lambda=0$ for consistency with
two remaining equations (11), (12). Integrating from (13) using (9)
we have
\begin{eqnarray}
f_{1}(t)=f_{1}(0)exp(\pm\frac{3}{4}\sqrt{\frac{\Lambda}{1-2\lambda}}t)
\end{eqnarray}
Also for satisfying (11), (12) we must fix
\begin{eqnarray}
f_{3}(t)=0
\end{eqnarray}
Finally comparing with (11), (12) we obtain the next solution under
constraint$\Lambda=0$. Indeed by adding (11), (12) we have
\begin{eqnarray}
(3\lambda-1)\ddot{u}+\frac{3}{2}\Lambda+\frac{1}{4}(1+2\lambda)\dot{u}^2=0
\end{eqnarray}
Substituting (9) with (15,16) in (17) we arrive to
\begin{eqnarray}
\hbox{either} \quad \Lambda = 0 \quad \hbox{or} \quad \lambda=
\frac{7}{10}.
\end{eqnarray}
Remember that $ \lambda=\frac{7}{10}$ is in contradiction with (14).
Thus the only possible choose is $\Lambda=0$. Consequently from (9)
we have
\begin{eqnarray}
 u(x,t)=\frac{4}{3}\log(x)+c'
  \end{eqnarray}
Which is obviously a no caustic solution,since $K=u_{t}=0$ and does
not diverge at all. Equation (19) does not represent a caustic but
it is certainly odd (in many ways its worse than a caustic). After a
shift in coordinates to eliminate $c'$, this corresponds to the
3-metric
\begin{eqnarray}\nonumber
g_{ij}=diag(1,x^{\frac{4}{3}},x^{\frac{4}{3}})
 \end{eqnarray}
Thus the spatial metric exhibits a naked singularity.

\subsection{$R\neq0$ solution for field equations in IR limit}
In this section we investigate the case with non vanishing Ricci
curvature. By elimination of the $\dot{u},u_{x}$ from  equations
(10)-(12) which makes  them non linear we relieve

  \begin{eqnarray}
  \frac{u_{tt}}{c_{eff}^{2}}=u_{xx}+\frac{2\Lambda(2\lambda+1)}{2\lambda-3}
  \end{eqnarray}
  where in it
  \begin{eqnarray}
  c_{eff}^2=\frac{-5+8\lambda+4\lambda^{2}-4\Lambda+8\lambda\Lambda}{2
\lambda-3}
  \end{eqnarray}
  since we want that the potential function $u(x,t)$ be stable  and
  with a  real sound speed, we must have
  \begin{eqnarray}
  \frac{-5+8\lambda+4\lambda^{2}-4\Lambda+8\lambda\Lambda}{2\lambda-3}>0
  \end{eqnarray}
  or equivalently
  \begin{eqnarray}
\hbox{either} \quad \lambda<\frac{1}{2} \quad \hbox{or} \quad
\lambda>\frac{3}{2}.
   \end{eqnarray}
    thus the general form of the metric function
  $u(x,t)$ is nothing but the familiar d'Alembert solution
  \begin{eqnarray}
  u(x,t)=H(x\pm c_{eff}t)-\frac{\Lambda(2\lambda+1)}{2\lambda-3}x^{2}
  \end{eqnarray}
The corresponding spatial metric (introducing an arbitrary function
$H(\xi)$ and a distance sacele a) is
\begin{eqnarray}\nonumber
g_{ij}=diag(1,H(x\pm c_{eff}t)e^{-ax^2},H(x\pm c_{eff}t)e^{-ax^2})
\end{eqnarray}
This is an arbitrary moving pulse of geometry on a Gaussian
background. We mention here that the cosmological constant term
treats as a
  forcing term and if we suppose that the metric functions been bound
  functions in some fixed points in the gravitational part of the
  manifold, then it seems that at an instant of time calling it $t=0$
  the system is under some initial force. Since we hope that the wavy
  solutions of the gravitational fields far from it's origin treat as
  classical GR,  thus we can say that in HL theory the
  cosmological constant term behaves as an external fixed force field.
    \section{Searching for Caustic singularity in UV regime and
$\lambda\neq1$}
  In this section we examine the existence of caustic singularity
  beyond the IR limit, so we add a non linear term to action (1). For
  $z=3$ the minimal coupling must be as a quadratic term of
  curvature. Thus we write the full modified action as
  \begin{eqnarray}
  I=\frac{M_{Pl}^{2}}{2}\int
  Ndt\sqrt{g}d^{3}\overrightarrow{x}(K^{ij}K_{ij}-\lambda
  K^{2}+R-2\Lambda+\xi R^2)
  \end{eqnarray}
  If we find the exact solutions for metric function
  $\partial_{t}u(x,t)$ does not diverge, then we determine that the non
  linear terms success in stoppage of formation such singularities. For
  ansatz (6) and action (25), it is easy to show that the reduced
  equation of motion for metric function $u(x,t)$ is
  \begin{eqnarray}
  (\frac{1}{2}+\lambda)u_t^2+u_{xx}(1-6\xi u_{xx}+\frac{19}{2}\xi
  u_x^2)+u_x^2(\frac{1}{2}-\frac{3}{4}\xi u_x^2)-8\xi
  u_xu_{xxx}-8\xi u_{xxxx}=u_{tt}
  \end{eqnarray}
  This equation is similar to a Hamilton-Jacobi equation. Then the
  general solution can be written as an additive function
  \begin{eqnarray}
  u(x,t)=f(x)+h(t)
    \end{eqnarray}
    Since we simultaneously must have the Ricci flat condition (9)
    thus the only possible consistent form is
 \begin{eqnarray}
  u(x,t)=\frac{4}{3}\log(f_{1}(t))+\frac{4}{3}\log(x)+c
    \end{eqnarray}

  Substituting this ansatz in (26)
 we obtain
  \begin{eqnarray}
   h_{tt}=\frac{C}{1-2\lambda}+\frac{(2\lambda-1)h_{t}^2-4\Lambda}{2(1-2\lambda)},
  \end{eqnarray}
where $C$ is insignificance constant.
  Remember that $K=u_{t}=h_{t}$ and we are searching for such solution
  that $k\rightarrow\infty$. The general solution for (29) is nothing
  else as an
  elementary function:
  \begin{eqnarray}
  h(t)=
  -\frac{t\sqrt{4\Lambda-2C}}{\sqrt{2\lambda-1}}-b+2\log(-c_{2}+c_{1}e^{\frac{t\sqrt{4\Lambda-2C}}{\sqrt{2\lambda-1}}})
  \end{eqnarray}
  ($C,c_1,c_2$ all are constant as a function of $\lambda,\Lambda$). The extrinsic
  curvature is
  \begin{eqnarray}
  K=\sqrt{\frac{4\Lambda-2C}{2\lambda-1}}\frac{c_{2}+c_{1}e^{\frac{t\sqrt{4\Lambda-2C}}{\sqrt{2\lambda-1}}}}
  {-c_{2}+c_{1}e^{\frac{t\sqrt{4\Lambda-2C}}{\sqrt{2\lambda-1}}}}
  \end{eqnarray}
  Obviously in contrast  of the former Mukohyama's hypothesis, it seems that this
  exact solution stricken
   to the caustic singularity. This would diverges as
  \begin{eqnarray}
    t\rightarrow
   t_{c}= \sqrt{\frac{2\lambda-1}{4\Lambda-2C}}\ln(\frac{c_{2}}{c_{1}})
  \end{eqnarray}
But  the $x$ dependence part of (28) i.e $\frac{4}{3}\log(x)+c$ does
not satisfy the related $x$ equation (as we will show). Thus this
solution which causes a harmful caustic singularity is not
acceptable. Thus although in UV regime this kind of singularity is
absent at least in order of $R^2$ action. Indeed (26) with (27)
ansatz leads to the following non linear fourth oreder differential
equation for $f(x)$:
 \begin{eqnarray}
   f_{xxxx}+\frac{1}{32\xi}(-38\xi
   f_{xx}f_{x}^2-4f_{xx}-2f_{x}^2+24\xi f_{xx}^2+3\xi f_{x}^4+32\xi
   f_{x}f_{xxx})=0
  \end{eqnarray}
As we know \cite{Murphy}this is a  4th order ordinary differential
equation (ODE) that is missing the dependent variable $x$. The order
can be reduced by introducing a new variable $p(x) = y_{x}$. If the
reduced ODE can be solved for p(x), the solution to the original ODE
is determined as a quadrature.The transformation
\begin{eqnarray}
y_{x}=p(x),y_{xx}=p(x)p(x)_{x},..
\end{eqnarray}
yields a reduction of order. If the reduced ODE can be solved for
$p(y)$, the solution to the original ODE can be given implicitly.\\
It is so easy to show that the function $\frac{4}{3}\log(x)+c$ is
not a solution for this ODE. Thus the caustic singularity is not
formed in UV regime as IR limit.  This would not be caustic, the
system does not remains in the IR regime insofaras the $\lambda$
should deviate from 1 by RG flow. These simple calculations prove
that there is no
   caustic singularity in plane symmetry even when the non linear
   curvature terms insert in the action. Consequently the former Mukohyama's conjecture \cite{Caustic}
    about
the avoidance of caustic singularity in HL theory is true.
  \section{Conclusion}
  In this short letter we show that the  Mukohyama's conjecture about
the avoidance of caustic singularity in HL theory is true. First we
showed that in IR limit, there is a family of exact solutions with
no caustic singularity and further we showed that the non Ricci flat
solutions define an effective sound speed. The stability test  of
this solution  under perturbations  show that there is a bound for
the value of $\lambda$. Secondly we show that even in non linear
regime, i.e. UV region the caustic singularity does not produce in a
satisfactory scheme. That solution which causes such singularity
does not satisfy fields equations correctly and completely. Thus our
work may be an adhoc analytic proof for former Mukohyama's
conjecture about the avoidance of caustic singularity in HL theory.

  \section{Acknowledgement}
  The authors thank from Anzhong Wang from Baylor (USA)for suggesting
  the topic of research. The work of M. R. Setare has
  been supported by Research Institute for Astronomy and Astrophysics
  of Maragha, Iran.

\end{document}